\documentclass[dvips]{article}
\usepackage{amstex}
\usepackage{epsfig}

\newcommand{\ket}[1]{\ensuremath{|#1\rangle}}
\newcommand{\xket}[1]{\ensuremath{\otimes\ket{#1}}}
\newcommand{\iket}[1]{\ensuremath{\ket{#1}_{\mathrm{i}}}}
\newcommand{\oket}[1]{\ensuremath{\ket{#1}_{\mathrm{o}}}}
\newcommand{\mtx}[1]{\ensuremath{\mathsf{#1}}}
\newcommand{\ftm}[1]{\ensuremath{\Bbb{F}_{#1}}}
\newcommand{\mtxspace}{\ensuremath{\mathcal{M}(M,N)}}

\begin{document}
%\baselineskip=2\baselineskip

\author{Giovanni Costantini and Fabrizio Smeraldi \\
\\
\textit{Dipartimento di Fisica dell'Universit\`a di Genova,}\\
\textit{via Dodecanneso 33, 16146 Genova, Italia}
}
\title{A Generalization of Deutsch's Example}
\date{}           
\maketitle

\begin{abstract}
Quantum parallelism is the main feature of quantum computation. 
In~1985 D.\@~Deutsch showed that a single quantum computation may be
sufficient to state whether a two--valued function of a two--valued variable is
constant or not. Though the generalized problem with unconstrained 
domain and range size admits no deterministic quantum solution, 
a fully probabilistic quantum algorithm is presented
in which quantum parallelism is harnessed to achieve a quicker 
exploration of the domain with respect to the classical ``sampling'' strategy.
\end{abstract}

\section{Introduction}
A quantum computer is a quantum system whose time evolution can be thought
of as a computation, much in the same way as we think of the time evolution
of a pocket calculator to be a computation.

For our pourposes it will suffice to model the quantum system as a ``black
box'' and focus our attention on two discrete observables out of a complete
set, which we shall call the input and output register. Following
the standard notation~\cite{Ekert_Jozsa}, we shall indicate
the computation of a function $f:A\to B$ as
\begin{equation*}
\ket{x} \xket{0} \mapsto \ket{x}\xket{f(x)},
\end{equation*}
the first ket describing the state of the input register and the second the
state of the output register. Kets are labelled according to the elements
of $A$ and $B$ they represent.

One of the most powerful features of quantum
computation is \emph{quantum parallelism\/}. The superposition principle of
quantum mechanics allows us to prepare the computer in a coherent
superposition of a set $I\subseteq A$ of input states. After a single run, 
all of
the corresponding outputs $f(x)$ appear in the final state, according 
to the time evolution 
\begin{equation}
\label{TheStandardComputation}
\sum_{x\in I}|x\rangle \otimes |0\rangle \mapsto \sum_{x\in I}|x\rangle
\otimes |f(x)\rangle
\end{equation}
Unfortunately, this is no ``pay one, take N''. In fact, the result is an
entangled state of the input and output registers and there is no
single measurement allowing us to extract from it all the computed 
values of $f(x)$~\cite{Jozsa}.
However, it may well be possible to distil from this final state some 
\emph{global property}
of the function, thus exploiting quantum parallelism. One of
the most famous examples was presented by D.\@~Deutsch, who showed that
a single quantum computation may suffice to state whether a two--valued
function of a two--valued variable is constant or not~\cite{Jozsa,Deutsch}. 
D.\@~Deutsch and R.\@~Jozsa later generalized this result~\cite{Deutsch_Jozsa} 
showing that the problem of classifying a given function
\begin{equation*}
f:\{0,\ldots,2N-1\}\to\{0,1\}
\end{equation*}
as ``not constant'' or ``not balanced'' can be solved in polynomial time
by means of a quantum computer (the time required by a classical solution
is exponential). Also D.\@~R.\@~Simon showed that the problem of determining if
a function is invariant under a so--called XOR mask, while it is classically
intractable, admits an efficient quantum solution~\cite{Simon}.

      All of the algorithms cited above (apart from the last, for which 
Simon also considered a fully probabilistic generalization) are 
characterized by a variable running time and zero error probability.
They consist of a non--classical computation like~(\ref{TheStandardComputation})
followed by a measurement of the final state of the computer, as a result
of which either the correct answer is obtained or the relevant information
is destroyed and an explicitly inconclusive result is returned. In the latter
case one has to go through the whole procedure again, so that only an average
running time for the algorithm can be estimated.

      Global properties of 
functions that can be determined by such an algorithm are said to be
Computable by Quantum Parallelism (QPC).This definition was put forward by 
Jozsa~\cite{Jozsa} who also
demonstrated that, at least in the case of two--valued functions, the QPC 
properties that can be determined by means of a single computation
are an exponentially small fraction of all the possible global properties. 

        In this paper we tackle the general problem of stating whether a 
function
\begin{equation*}
f:\{0,\ldots,N-1\} \to \{0,\ldots,M-1\}
\end{equation*}
is constant or not. 

      We show that for $N>2$ or $M>2$ this property is not QPC, 
meaning that any measurement following a computation like~%
(\ref{TheStandardComputation}) has a finite probability of yielding a wrong
result. We therefore investigate the power of quantum parallelism in a
fully probabilistic setting.
Assuming that the (classical) computation of $f$ on $k$ 
randomly sampled points yields a constant value, we calculate the posterior
probability that $f$ is actually constant. We then compute the analogous 
probability for a quantum algorithm requiring the same number of computations
of $f$. Comparison of the two results shows that our quantum strategy allows
making a better guess at the solution, its indications being more likely to
be correct.

\section{Is a direct generalization of Deutsch's example possible? 
\label{Impossibility}
\label{TheExample}}

      We shall now briefly recall the classical example put forward by
D.\@~Deutsch~\cite{Deutsch}, before confronting the problem of its
generalization.
Suppose we are given a function $f:\{0,1\}\to \{0,1\}$ and we are interested 
to know whether $f$ is constant or not. Of course
there are only four such functions (i.e. four instances of the problem),
namely 
\begin{equation}
\begin{array}{llll}
\left\{ 
\begin{array}{c}
f_1(0)=0 \\ 
f_1(1)=0
\end{array}
\right. & \left\{ 
\begin{array}{c}
f_2(0)=1 \\ 
f_2(1)=1
\end{array}
\right. & \left\{ 
\begin{array}{c}
f_3(0)=0 \\ 
f_3(1)=1
\end{array}
\right. & \left\{ 
\begin{array}{c}
f_4(0)=1 \\ 
f_4(1)=0
\end{array}
\right.
\end{array}
\label{TheFourFunctions}
\end{equation}

If all we can use is a classical computer, there is only one way to do the
job: we must compute \emph{both\/} $f(0)$ and $f(1)$ and compare
them to check if they are equal. 
On the contrary, since in this simple case the property ``$f$ is constant'' 
is QPC, a quantum computer gives us a fair chance of finding the
solution at the cost of the single computation
\begin{equation*}
{\frac 1{\sqrt{2}}}(|0\rangle +|1\rangle )\otimes |0\rangle \mapsto 
{\frac 1{%
\sqrt{2}}}(|0\rangle |f(0)\rangle +|1\rangle |f(1)\rangle ).
\end{equation*}

After the computation, the calculator halts  with its input and output
registers in one of four possible states, corresponding to the four possible
functions:
\begin{equation}
\label{FinalStates}
\begin{aligned}
|f_1\rangle &={\frac 1{\sqrt{2}}}(|0\rangle |0\rangle +|1\rangle |0\rangle ) \\
|f_3\rangle &={\frac 1{\sqrt{2}}}(|0\rangle |0\rangle +|1\rangle |1\rangle ) 
\end{aligned} \qquad
\begin{aligned}
|f_2\rangle &={\frac 1{\sqrt{2}}}(|0\rangle |1\rangle +|1\rangle |1\rangle ) \\
|f_4\rangle &={\frac 1{\sqrt{2}}}(|0\rangle |1\rangle +|1\rangle |0\rangle )
\end{aligned}
\end{equation}

Since the above states are linearly dependent,
they cannot be distinguished with certainty. This means that no measurement
can establish which function was actually computed, or, which is the same,
it's impossible to extract from the final state \emph{both\/} the
values of $f(0)$ and $f(1)$. However, we need only discriminate 
$|f_1\rangle $ and $|f_2\rangle $,
the final states yielded by the constant functions, from $|f_3\rangle $ and 
$|f_4\rangle $. This can actually be done by measuring on the final state of
the two registers an observable with the following non--degenerate
eigenstates: 
\begin{equation*}
\begin{aligned}
|\mathrm{SAME}\rangle &={\frac 12}(|0\rangle +|1\rangle )
(|0\rangle -|1\rangle ) \\
|\mathrm{FAIL}\rangle &={\frac 12}(|0\rangle +|1\rangle )
(|0\rangle +|1\rangle ) 
\end{aligned} \quad 
\begin{aligned}
|\mathrm{DIFFERENT}\rangle&={\frac 12}(|0\rangle -|1\rangle )
(|0\rangle -|1\rangle ) \\
|\mathrm{ERROR}\rangle &={\frac 12}(|0\rangle -|1\rangle )
(|0\rangle +|1\rangle )
\end{aligned}
\end{equation*}
These four states can be thought of as ``flags'' indicating the result of
the computation and have been named according to their meaning. 
This becomes clearer as soon as we rewrite the final states~(\ref{FinalStates})
on the basis of the above eigenvectors: 
\begin{equation*}
\begin{aligned}
|f_1\rangle &={\frac 1{\sqrt{2}}}(\ket{\mathrm{FAIL}}+\ket{\mathrm{SAME}}) \\
|f_2\rangle &={\frac 1{\sqrt{2}}}(|\mathrm{FAIL}\rangle -|\mathrm{SAME}%
\rangle )
\end{aligned} \qquad
\begin{aligned}
|f_3\rangle &={\frac 1{\sqrt{2}}}(|\mathrm{FAIL}\rangle +|\mathrm{DIFFERENT}%
\rangle ) \\
|f_4\rangle &={\frac 1{\sqrt{2}}}(|\mathrm{FAIL}\rangle 
-|\mathrm{DIFFERENT}\rangle )
\end{aligned}
\end{equation*}

It is now evident that
\begin{enumerate}
\item  projection along the eigenvector $|\mathrm{SAME\rangle }$ can only
take place if the state of the computer is either $|f_1\rangle $ or $%
|f_2\rangle $, i.e.\@ when $f$ is constant;
\item  likewise, projection along the eigenvector $|\mathrm{DIFFERENT\rangle 
}$ can only occur if $f$ is not a constant function;
\item  regardless of the final state $|f_i\rangle $ of the two registers
after the computation, the measurement can yield a $|\mathrm{FAIL\rangle }$
state with probability $1/2$;
\item  state $|\mathrm{ERROR}\rangle$ is orthogonal to the four final
states listed above, and therefore it should show up only as
a consequence of noise--induced
errors.
\end{enumerate}

Note that, according to the definition  of the QPC class~\cite{Jozsa},
the quantum algorithm can either give us the
correct answer or no answer at all: as long as everything works properly,
we'll never get a wrong result. This comes in handy when we are asked to
solve a decision--theoretic problem in which simply waiting has a much 
higher utility than taking a wrong action. In this case we can  
discard the FAIL results and base our decisions upon the meaningful 
answers, which we know to be correct.

The straightforward generalization of Deutsch's example would go as follows.
Given a function
\begin{equation}
\label{f}f:\{0,1,\ldots,N-1\}\to\{0,1,\ldots,M-1\}
\end{equation}
and assuming we can perform the non--classical computation
\begin{equation}
\label{TheComputation}
\frac{1}{\sqrt{N}}
\left(\sum_{x=0}^{N-1}\ket{x}\right)\xket{0}\mapsto
\frac{1}{\sqrt{N}}\sum_{x=0}^{N-1}\ket{x}\xket{f(x)},
\end{equation}
we are asked to  devise an observable $\mathcal{O}$ on the joint state of the 
two registers
such that, after a single measurement of $\mathcal{O}$, we can either
\begin{enumerate}
\item{} obtain a reliable indication that  function $f$ is constant;
\item{} obtain an equally reliable indication that $f$ is not constant, or
finally
\item{} get an explicitly inconclusive result.
\end{enumerate}

Let $\mathcal{H}$ be the Hilbert space of the joint states of the
input and output registers.
If $\mathcal{E}$ is the basis of $\mathcal{H}$ 
formed by the (non--degenerate)
eigenstates of $\mathcal{O}$, all that is needed would be the
existence of two disjoint subsets $\mathcal{C},\overline{\mathcal{C}}
\subset\mathcal{E}$ such that
\begin{enumerate}
\item[\textit{i.}] all the final states obtained from the computation of  
constant (non--con\-stant) functions have 
a non--zero projection along $\mathcal{C}$ (along $\overline{\mathcal{C}}$);
\item[\textit{ii.}] the final states corresponding to non--con\-stant 
(con\-stant) functions are orthogonal
to $\mathcal{C}$ (to $\overline{\mathcal{C}}$).
\end{enumerate}
These two requirements are evidently 
fulfilled in the case of Deutsch's  example, as can be easily seen by taking
$ \mathcal{C}=\left\{\ket{\mathrm{SAME}}\right\}$ and
$\overline{\mathcal{C}}=\left\{\ket{\mathrm{DIFFERENT}}\right\}$
(for further details see~\cite{Jozsa}). 

However, as soon as the domain
and range of the function $f$ grow larger, requirements \textit{i.\@} and 
\textit{ii.\@} become incompatible.
What happens is
that whenever $N>2$ or~$M>2$ the computation of constant functions yields
final states that are linearly dependent upon those obtainable from
non--constant functions. This clearly forbids the existence of 
$\mathcal{C}$, since the final states coming from non--constant functions
cannot be orthogonal to $\mathcal{C}$.

      In other words, the global property ``$f$ is constant'' is no longer 
QPC in the general case. Note that, as demonstrated by Jozsa~\cite{Jozsa}, 
this result is
independent of the particular superposition used
as the input state for the non--classical computation~(\ref{TheComputation}).  

\section{Probabilistic generalization\label{Generalization}}

      The fact that the investigated property of $f$ is not
QPC compels us to work in a fully probabilistic setting in order to cope with 
the possibility of wrong results.
Preserving the general structure of the algorithm as outlined at the
beginning of the preceding section, we note that we can still
devise an observable $\mathcal{O'}$ such that
any ``\mbox{con}\-\mbox{stant}'' ket has a large projection along 
a subset $\mathcal{C'}$
of the eigenstates of $\mathcal{O'}$; for example, we can arrange for
$\mathrm{Span}(\mathcal{C'})$
to be the very space spanned by the ``constant'' vectors. The problem is now 
that since ``non--constant'' kets generally have a non--zero
projection along $\mathcal{C'}$, measuring $\mathcal{O'}$ no longer ensures 
a clear--cut distinction between constant and non--constant functions.
However, since ``non--constant'' final states do have some component along the
orthocomplement of $\mathcal{C'}$, measuring $\mathcal{O'}$ still gives
some (probabilistic) information about the computed function $f$.
We are left with two asymmetrical possibilities (actually, as
we shall see, a more convienient choice for $\mathcal{C'}$ also  
makes an explicitly inconclusive result possible):
\begin{enumerate}
\item[\textit{a.}] measuring $\mathcal{O'}$ yields an eigenvalue associated 
to the orthocomplement of $\mathrm{Span}(\mathcal{C'})$ in
$\mathcal{H}$. Since this can only happen if the 
computed function $f$ is
\emph{not\/} constant, this is an exact solution to the problem.
\item[\textit{b.}] measuring $\mathcal{O'}$ projects the final state of 
the two registers onto a state in $\mathcal{C'}$. 
If the computed
function were constant, this would be the only possibility; unfortunately,
as seen above, other functions may also yield the same result. We
have therefore obtained only a probabilistic indication about $f$ being
constant.
\end{enumerate}

It is now clear that the generalized algorithm is essentially similar
to a classical probabilistic algorithm, in that its results are not
necessarily correct. Nevertheless, as we shall see in the following
sections, the posterior probability of the function actually being
constant after a result of type \textit{b.\@} is obtained
turns out to be much larger for our quantum algorithm than for the 
classical ``sampling'' strategy (see section~\ref{Evaluation}).

In the rest of this section we shall deal with the choice of the observable,
which constitutes the core of the algorithm.

\subsection{Functions and matrices\label{FunctionsAndMatrices}}

We would now like to introduce a correspondence between the Hilbert
space $\mathcal{H}$ of the two registers of the computer and
the space $\mtxspace$ of complex matrices with $M$ rows
and $N$ columns.

Let $\mathcal{B}$ 
be the computational basis of $\mathcal{H}$, the first ket referring to the
state of the input register and the second to that of the output register:
\begin{equation*}
\mathcal{B}=\{\iket{0}\oket{0},\iket{0}\oket{1},\ldots,
\iket{N-1}\oket{M-1}\}
\end{equation*}
We define the isomorphism $\varphi:\mathcal{H}\rightarrow\mtxspace$
by identifying $\varphi\left(\iket{n}\oket{m}\right)$ with the $M\times N$
matrix whose elements are
\begin{equation*} 
\left[\varphi\left(\iket{n}\oket{m}\right)\right]_{i,j}=
\delta_{m,i}\delta_{n,j}.
\end{equation*}
The isomorphism $\varphi$, which maps the
elements of $\mathcal{B}$ onto the canonical basis of \mtxspace, is then 
extended by linearity to the whole $\mathcal{H}$.

Since the final state of the computer after the computation of function
$f$ is
\begin{equation*}
\ket{f}=\frac{1}{\sqrt{N}}\sum_{i=0}^{N-1}\iket{n}\oket{f(n)},
\end{equation*}
the entries of the corresponding matrix $\mtx{F}=\varphi(\ket{f})$ turn
out to be $(\mtx{F})_{n,m}=\delta_{m,f(n)}$,
so that \mtx{F} somehow resembles the graph of $f$ drawn with the ``$x$'' axis
along the rows and the ``$y$'' axis pointing down
\footnote{For convenience we write $\mtx{F}_{0,0}$ instead of $\mtx{F}_{1,1}$
for the upper left element of matrix \mtx{F}.}.

      It is easy to check that the scalar product
\begin{equation}
\mathsf{F}\cdot\mathsf{G}:=\mathrm{Tr}\left(\mathsf{F}^{\dagger}
\mathsf{G}\right)
\label{TheDotProduct}
\end{equation}
in \mtxspace\ is preserved by $\varphi$, i.e.\@~$\langle f|g\rangle=
\mtx{F}\cdot\mtx{G}$ for any two vectors \ket{f}, \ket{g} of $\mathcal{H}$
(we write $\langle\cdot|\cdot\rangle$ for the scalar product in 
$\mathcal{H}$.)

\subsection{The Fourier Transform Matrices basis\label{TheFTMBasis}}

We shall now construct the observable $\mathcal{O'}$ 
as specified at the beginning of this section.

An observable in ${\cal H}$ is identified by its $MN$ eigenstates that form
an orthogonal basis, or, using the isomorphism $\varphi$, by $MN$ orthogonal
matrices in \mtxspace.
We propose to take the $MN$ two--parameter matrices 
\ftm{\alpha ,\beta } with $\alpha =0,1,\ldots ,M-1$ and 
$\beta =0,1,\ldots ,N-1$ whose entries are defined by 
\begin{equation}
\label{Fourier}\left( \ftm{\alpha ,\beta }\right) _{m,n}=\frac 1{\sqrt{MN}%
}\,e^{i\frac{2\pi }M\alpha m}e^{i\frac{2\pi }N\beta n}\text{.} 
\end{equation}

We shall call the above matrices \emph{Fourier Transform Matrices} (FTM).
We recall that given a matrix $\mtx{A}\in \mtxspace$, its 
two dimensional Discrete Fourier Transform 
$\widetilde{\mtx{A}}\in \mtxspace$ is defined as 
\begin{equation*}
\left(\widetilde{\mtx{A}}\right)_{\alpha,\beta}=
\frac 1{\sqrt{MN}}\sum_{m,n}\left(\mtx{A}\right)_{m,n}
e^{-i\frac{2\pi }M\alpha m}e^{-i\frac{2\pi }N\beta n} 
=\mathrm{Tr}\left(\ftm{\alpha,\beta}^{\dagger }\mtx{A}\right) =
\ftm{\alpha,\beta}\cdot \mtx{A}\text{.} 
\end{equation*}
Therefore the components of \mtx{A} on the FTM basis are the entries 
of its Discrete Fourier Transform $\widetilde{\mtx{A}}$.

We still have to
decide which eigenvectors are to be taken as an indication of the
function~$f$ being constant. In other words we have
fixed the basis ${\cal E}$ but have yet to choose the subset ${\cal C'}$. 
We take ${\cal C}^{^{\prime }}$ as composed by the $M-1$ matrices 
\ftm{\alpha ,0}, with $\alpha =1,2,\ldots ,M-1$. It is easy to check that
${\cal C}^{^{\prime }}\cup\left\{\ftm{0,0}\right\}$ spans the subspace of
\mtxspace\ generated by the set of the $M$ matrices 
$\left\{\varphi(\ket{f})|f\mathrm{\ constant}\right\}$ corresponding to
the constant functions. We have not included 
\ftm{0,0} in $\mathcal{C}^{\prime}$ because the projection probability of the
computer's final state on \ftm{0,0} is the same for all functions:
\begin{equation}
\label{PrFAIL}
\ftm{0,0}\cdot \mtx{F}=
\mathrm{Tr}\left(\ftm{0,0}^\dagger\mtx{F} \right)=\frac{1}{\sqrt{M}}
\qquad\forall\mtx{F}\in\mtxspace.
\end{equation}
Therefore \ftm{0,0} has the same role that state \ket{\mathrm{FAIL}}
had in Deutsch's example. In the following we shall put 
$\varphi^{-1} \left(\ftm{0,0} \right)=\ket{\mathrm{FAIL}}$ 
and we shall speak equiv\-al\-ently of the matrix 
\ftm{0,0} in \mtxspace \ or of the state \ket{\mathrm{FAIL}} in ${\cal H}$.

Likewise, since $\ftm{0,\beta}\cdot\mtx{F}=0$ for all
$\beta\neq 0$ and every matrix \mtx{F}, subset
\begin{equation*}
\mathcal{E}_{error}=\{\ftm{0,\beta}|1\le\beta\le N-1\}
\end{equation*}
plays the role of the \ket{\mathrm{ERROR}} state in Deutsch's example 
(section~\ref{TheExample}). 

      The remaining FTM matrices constitute set $\overline{\mathcal{C}}$:
\begin{equation*}
\overline{\mathcal{C}}=
\{\ftm{\alpha,\beta}|1\le\alpha\le M-1, 1\le\beta\le N-1\}.
\end{equation*}
Note that we did not put a prime on $\overline{\mathcal{C}}$, since it does
satisfy both conditions \textit{i.\@} and \textit{ii.\@} listed in section~%
\ref{Impossibility}. This accounts for the lack of symmetry we pointed out
at the beginning of section~\ref{Generalization}.

\section{
\label{Evaluation}
Efficiency of the probabilistic generalization}

      Suppose we run the quantum algorithm $k$ times on the same function $f$
and we always get an indication that $f$ is constant (a projection
onto $\mathcal{C'}$). We need to gauge the 
reliability of this result, which we can do by computing the posterior 
probability 
\begin{equation}
\label{prob}\Pr (\text{really constant}|\ k\text{``constant'' outcomes}) 
\end{equation}
($\Pr (\mathrm{const}|\ k)$ for short) that the function really is constant.
This quantity can also be used to compare the efficiency of the quantum 
algorithm against a conventional classical solution, since what we are
looking for is a procedure giving the lowest probability of error in change
for the same computational effort. 
 
To evaluate (\ref{prob}) we use Bayes' theorem, that is 
\begin{equation}
\label{bayes}\Pr (\text{const}|\ k)=\frac{\Pr (\text{const}\wedge k)}{\Pr (k)%
}\text{,} 
\end{equation}
where by $\Pr(\text{const}\wedge k)$ we mean the joint probability that 
$f$ is constant \emph{and\/} that $k$ runs of the algorithm yield a 
``constant'' outcome,  corresponding to the final state being projected
along $\mathcal{C}^{\prime}$. By the product rule, this can be expressed
as
\begin{equation}
\label{qnum}\Pr (\text{const}\wedge k)=\Pr (\text{const})\Pr (k|\text{const}%
). 
\end{equation}
Assuming a uniform probability distribution on all the
possible functions, we have $\Pr (\text{const})=M/M^N$. Regardless of the
computed function $f$, FAIL results have a $1/M$ probability of showing
up (see equation~\ref{PrFAIL}). This leads to   
$\Pr (k|\text{const})=\left( 1-1/M\right) ^k$. 
As a consequence~(\ref{qnum}) becomes
\begin{equation}
\label{num}\Pr (\text{const}\wedge k)=M^{1-N}\left( 1-\frac 1M\right) ^k. 
\end{equation}

The denominator of~(\ref{bayes}) can be expanded over all the $M^N$ 
possible functions of type~(\ref{f}): 
\begin{equation}
\label{total}\Pr (k)=\sum_{i=1}^{M^N}\Pr (f_i)\Pr (k|f_i). 
\end{equation}
Since we assumed the input functions to be uniformly distributed, we have
\begin{equation}
\label{unif}\Pr (f_i)=M^{-N}\quad \forall \ i\text{.} 
\end{equation}
The $k$ runs of the quantum algorithm are stochastically independent and
that implies that the likelihoods $\Pr (k|f)$ appearing in~(\ref{total})
are simply given by
\begin{equation}
\label{indip}\Pr (k|f)=\left[ \Pr ({\cal C}^{^{\prime }}|f)\right] ^k, 
\end{equation}
where with $\Pr ({\cal C}^{^{\prime }}|f)$ we mean the likelihood of a
single run, the probability of a projection onto ${\cal C}^{^{\prime }}$
when the function is $f$. So we can concentrate only on $\Pr ({\cal C}%
^{^{\prime }}|f)$, which, with the help of the sum rule, can be
expressed as
\begin{equation}
\label{key}\Pr ({\cal C}^{^{\prime }}|f)=\Pr (\frak{K}|f)-\Pr (\text{FAIL}%
|f) 
\label{totlik}=\sum_{\alpha =0}^{M-1}\Pr (\frak{K}_\alpha |f)-
\Pr (\text{FAIL}|f)\text{.} 
\end{equation}  
Here $\frak{K}$ stays for event ``after the measure the computer's final
state projects itself onto the subspace of all constant functions'', 
$\frak{K}_\alpha $ for the projection onto the matrix $\mtx{K}_\alpha $,
which represents the $\alpha$-th constant function and is defined as
$\left(\mtx{K}_\alpha\right)_{m,n}=\delta_{m,\alpha}$, and FAIL for the
projection onto $\left| \text{FAIL}\right\rangle $. We have used the
fact that, thanks to the orthogonality relations, events 
$\frak{K}_0,\frak{K}_1,\ldots,\frak{K}_{M-1}$
are mutually exclusive and so are ${\cal C}^{^{\prime }}$
and FAIL and that 
${\cal C}^{^{\prime }}\cup\mathrm{FAIL}=\frak{K}$ and
$\bigcup_{\alpha =0}^{M-1}\frak{K}_\alpha=\frak{K}$. 

As we have seen in~(\ref{PrFAIL}),  
$\Pr \left( \text{FAIL}|f\right) =1/M\quad \forall ~f$.
On the other hand 
\begin{equation}
\label{ter}\Pr \left( \frak{K}_\alpha |f\right) =
\left| \mtx{K}_\alpha \cdot \mtx{F}\right|^2=
\left| \mathrm{Tr}\left( \mtx{K}_\alpha ^{\dagger }\mtx{F}\right)\right|^2
\text{,} 
\end{equation}
where 
\begin{equation}
\label{Af}\mtx{F}=\frac 1{\sqrt{N}}\left( 
\begin{array}{cccc}
f_{0,0} & f_{0,1} & \cdots & f_{0,N-1} \\ 
f_{1,0} & f_{1,1} & \cdots & f_{1,N-1} \\ 
\vdots & \vdots & \ddots & \vdots \\ 
f_{M-1,0} & f_{M-1,1} & \cdots & f_{M-1,N-1} 
\end{array}
\right) 
\end{equation}
is the matrix related through isomorphism $\varphi $ to the computer's final
state when the function is $f$ (Note that most of the elements of \mtx{F}
are zero, since $f_{m,n}=\delta _{m,f(n)}$).

Let us now compute explicitly the trace that appears in the 
r.h.s.\@~of~(\ref{ter}):
\begin{equation}
\label{trace}
\mathrm{Tr}\left(\mtx{K}_{\alpha}^{\dagger}\mtx{F}\right)=
\sum_{m,n}(\mtx{K}_{\alpha})^*_{m,n} (\mtx{F})_{m,n}=
\frac{1}{N}\sum_{m,n}\delta_{m,\alpha} f_{m,n}=
\frac{1}{N}\sum_{n} f_{\alpha,n}.
\end{equation}
Equation (\ref{ter}) then becomes 
\begin{equation}
\label{finalpr}\Pr \left(\frak{K}_\alpha |f\right) 
=\frac 1{N^2}\left| \sum_{n=0}^{N-1}f_{\alpha ,n}\right| ^2\text{.} 
\end{equation}

Equation (\ref{trace}) contains the sum of the elements appearing in the $%
\alpha $-th row of matrix (\ref{Af}). Since matrix \mtx{F} has 
a sole one in
any column, this sum is equivalent to the number of ones 
in the $\alpha$-th row of \mtx{F}. This gives us an idea for a smart 
classification of all the
possible functions appearing in~(\ref{total}): we associate with every
function an $N+1$--uple $\left( j_0,j_1,\ldots ,j_N\right) $, where $j_l$ is
the number of rows of its corresponding matrix \mtx{F} with $l$ ones and 
$N-l$ zeroes. 
Doing so
we can replace the sum over $i $ appearing in (\ref{total}) by a sum over
the $N+1$--uples $\left( j_0,j_1,\ldots ,j_N\right) $, with conditions
\begin{gather}
\label{zero}0\leq j_l\leq M\qquad \forall ~l, \\
\label{one}j_1+2j_2+\cdots +Nj_N=N, \\
\label{two}j_0+j_1+\cdots +j_N=M.
\end{gather}
Condition~(\ref{one}) expresses the requirement that the total number of
ones in matrix \mtx{F} is $N$ (or, since each column contains a sole one,
that \mtx{F} has $N$ columns), while~(\ref{two}) 
is equivalent to the condition
that \mtx{F} has $M$ rows. In the following, we shall indicate with 
${\cal I}$ the set of
the $N+1$--uples $\left( j_0,j_1,\ldots ,j_N\right) $ that
satisfy equations~(\ref{zero})--(\ref{two}).

Note that, since every $N+1$--uple corresponds to more than one function,
when summing over the $N+1$--uples we must use the right combinatorial
factors. These, for a fixed $N+1$--uple $\left( j_0,j_1,\ldots ,j_N\right) $,
are given by: 
\begin{equation}
\label{comb}C_{j_0,j_1,\ldots ,j_N}=\frac{N!}{\left( 0!\right) ^{j_0}\left(
1!\right) ^{j_1}\cdots \left( N!\right) ^{j_N}}\cdot \frac{M!}{%
j_0!j_1!\cdots j_N!},
\end{equation}
the first term corresponding to column permutations and the second to
row permutations.

We can now use equation~(\ref{finalpr}) together with this way of
classifying the functions to evaluate the total likelihood $\Pr \left( 
\frak{K}|f\right) $ that appears as the first term in
equation~(\ref{key}).
If $f_{j_0,j_1,\ldots ,j_N}$ stands for a function corresponding to the
$N+1$--uple $\left( j_0,j_1,\ldots ,j_N\right) $,%
\begin{equation*}
\Pr \left( \frak{K}|f_{j_0,j_1,\ldots ,j_N}\right) =\sum_{\alpha
=0}^{M-1}\Pr \left(\frak{K}_\alpha |f_{j_0,j_1,\ldots ,j_N}\right) =\frac
1{N^2}\sum_{l=0}^N\left( l\cdot j_l\right) ^2\text{.} 
\end{equation*}
Consequently~(\ref{totlik}) becomes%
\begin{equation*}
\Pr \left( {\cal C}^{^{\prime }}|f_{j_0,j_1,\ldots ,j_N}\right) =\frac
1{N^2}\sum_{l=0}^N\left( l\cdot j_l\right) ^2-\frac 1M 
\end{equation*}
and~(\ref{indip}) becomes in turn
\begin{equation*}
\Pr \left( k|f_{j_0,j_1,\ldots ,j_N}\right) =\left[ \frac
1{N^2}\sum_{l=0}^N\left( l\cdot j_l\right) ^2-\frac 1M\right] ^k\text{.} 
\end{equation*}

Now we can sum over all the possible $N+1$--uples with 
conditions~(\ref{zero})--%
(\ref{two}) and with the combinatorial factors~(\ref{comb}), obtaining the
expression of equation~(\ref{total}) in the quantum case:%
\begin{equation*}
\Pr (k)=\sum\limits\Sb \left( j_0,j_1,\ldots ,j_N\right) \in {\cal I}\endSb
\frac 1{M^N}\left[ \frac 1{N^2}\sum_{l=0}^N\left( l\cdot j_l\right) ^2-\frac
1M\right] ^kC_{j_0,j_1,\ldots ,j_N} 
\end{equation*}
(N.B. we have used the fact that $\Pr (f)=M^{-N}\quad \forall ~f$, since we
suppose a prior uniform probability distribution on all the functions).

Finally, using also equation~(\ref{num}) we can express the posterior
probability~(\ref{bayes}) in the quantum case as 
\begin{equation}
\label{quantum}
\Pr (\text{const}|k)=
\frac{M\left( 1-\frac 1M\right) ^k}
{\sum\limits\Sb \left( j_0,j_1,\ldots ,j_N\right) \in {\cal I}\endSb 
\left[\frac 1{N^2}\sum_{l=0}^N\left( l\cdot j_l\right) ^2-\frac 1M\right]^k
C_{j_0,j_1,\ldots ,j_N}}\text{.}
\end{equation}

      In the following section we will derive the corresponding expression 
for the classical case.

\section{Efficiency of the ``sampling'' algorithm}

There exists at least one obvious classical probabilistic algorithm that can
be used to spot constant functions. We can simply compute the value of 
\begin{equation*}
f:\left\{ 0,\ldots ,N-1\right\} \to \left\{
0,\ldots ,M-1\right\} 
\end{equation*}
on $k$ randomly chosen points of its domain and decide that $f$ is
constant if its restriction to the sampled points is. This procedure,
which we shall call the ``sampling algorithm'', evidently constitutes the
best possible classical strategy to solve the problem, since it uses up all
the information we can gain on $f$ by $k$ classical computations.

      In order to allow a direct comparison with the quantum algorithm, 
we have to find out what the posterior probabilities $\Pr(\mathrm{const}|k)$ 
are in this case. Starting again from Bayes' theorem, 
we can express the numerator of equation~(\ref{bayes}) as 
\begin{equation*}
\Pr (\text{const}\wedge k)=\Pr (\text{const})Pr(k|\text{const}%
)=\Pr (\text{const}), 
\end{equation*}
that by~(\ref{unif}) is equal to $M/M^N=M^{1-N}$ (note that in the classical
case $\Pr(k|\mathrm{const})=1$, since no FAIL results exist).

We must now evaluate the denominator of Bayes' formula, namely equation~(\ref
{total}). 
Choosing the $k$ inputs at random actually turns out to be inessential as
long as the functions are uniformly distributed: sampling the first $k$ 
points $0,1,...,k-1$ is just as good.
Let us therefore divide all the possible functions into two classes. The
first is made up by those for which at least the first $k$ values are
constant; they are $MM^{N-k}=M^{N-k+1}$. All the other functions belong to
the second class. As a consequence, the likelihoods that appear in the
r.h.s. of (\ref{total}) are simply given by%
\[
\Pr (k|f)=\left\{ 
\begin{array}{l}
1 
\text{ \quad if $f$ belongs to the first class} \\ 0\text{ \quad otherwise} 
\end{array}
\right. 
\]
Putting this expression in (\ref{total}), and recalling (\ref{unif}), we can
rewrite~(\ref{bayes}) as 
\begin{equation}
\label{bayes2}\Pr (\text{const}|k)=\frac{M^{1-N}}{_{M^{N-k+1}/M^N}}=M^{k-N}%
\text{.} 
\end{equation}

        This result is to be compared with equation~(\ref{quantum}), 
which gives the corresponding posterior probability
after $k$ runs of the quantum algorithm. In order to do so, formula~%
(\ref{quantum}) must evidently be evaluated by means of
a (classical!) computer. Before listing the numerical results, however, 
we are going to discuss two special cases that can be solved analytically
in the limit of large $N$.

\section{Worst case and best case analysis}

We shall now analyse the behaviour of our generalized quantum algorithm in
the worst possible case, that is when the computed function has maximum
probability of being mistaken for a constant function, even if it is not.
This occurs quite naturally for a matrix of the following kind:
\bigskip
\begin{equation*}
\mathsf{G}=
\frac{1}{\sqrt{N}}
\left.
\left(
\hspace{1 mm}
\smash[t]{
\overbrace{
        \begin{array}[c]{cccccc}
                1 & 1 & 1 & \ldots & 1 & 0 \\
                0 & 0 & 0 & \ldots & 0 & 1 \\
                0 & 0 & 0 & \ldots & 0 & 0 \\
                \vdots & \vdots & \vdots & \ddots & \vdots & \vdots \\
                0 & 0 & 0 & \ldots & 0 & 0 
        \end{array}
}^{N}
} 
\hspace{1 mm}
\right)
\right\}\scriptstyle{M}
\end{equation*}
representing a function $g$ that is constant on its whole domain but 
for one point.

      The resulting probability of error is given by the squared modulus of
the projection of \mtx{G} on the space spanned by the set 
$\left\{\mtx{K}_0,\ldots,\mtx{K}_{M-1}\right\}$ 
of the matrices associated to constant functions~%
\footnote{We assume that $M$ is so large that the probability of a FAIL
result is negligible.}, that is
\begin{equation*}
\Pr{}_{E}=1-\frac{2}{N}+\frac{2}{N^2}.
\end{equation*}
Therefore $\Pr{}_{E}$  tends to one in the limit of large $N$. 
Here again, in order to compensate for this we have to run the quantum 
algorithm several times,
say $k$ (classically, we would have to sample more and more points). If we
want to keep the probability of being ``cheated'' by an almost--constant
function $g$
as low as a given value $\varepsilon$, we evidently have to choose $k$ so that
$\left(\Pr{}_{E}\right)^{k}=\varepsilon$,
that is
\begin{equation*}
k=\frac{\ln\varepsilon}{\ln\left(1-\frac{2}{N}+\frac{2}{N^{2}}\right)}.
\end{equation*}

      As we would expect, $k$ does tend to infinity in the limit of large $N$,
meaning that exploring an even larger domain requires an infinite number
of computations. It is nevertheless interesting to study the ratio
$\eta=k/N$ of the number
of runs to the number of elements in the domain. In the limit of large $N$, 
this becomes
\begin{equation}
\label{WorstCaseErrorProbability}
\lim_{N\to\infty}\eta=\lim_{N\to\infty}\frac{k}{N}=
-\frac{\ln\varepsilon}{2},
\end{equation}
which is a constant independent on $N$.

      Therefore, if we are required to perform the computation with a worst
case error probability $\varepsilon$, we have to run our quantum computer a 
number
of times which, in the limit of large $N$, is a definite fraction $\eta$
of $N$.  Equation~(\ref{WorstCaseErrorProbability}) can in this case be 
inverted to
obtain $\varepsilon$ as a function of $\eta$, yielding 
$\varepsilon=e^{-2\eta}$.

      Coming now to the classical case, sampling a fraction $\eta$ of the
points in the domain (which requires $k=\eta N$ computations) entails having
a probability $\varepsilon = 1-\eta$ of mistaking $g$ for $f_0$.
\begin{figure}
      \epsfig{file=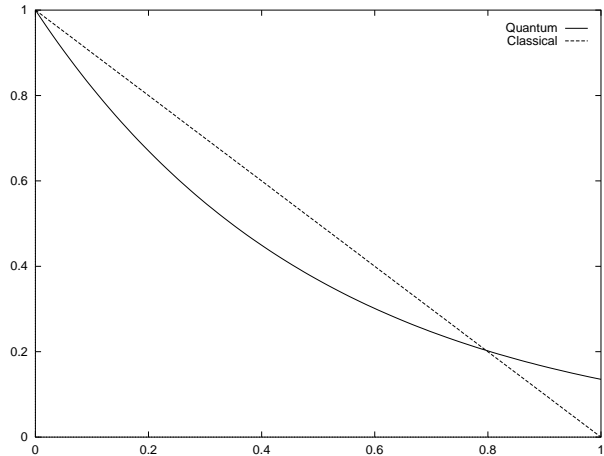, width=8.6cm}
      \caption{Worst case probability of error $\varepsilon$ versus
the ratio $\eta=k/N$ of the number of computations to the number of
points in the domain. Solid line: quantum algorithm; dashed line:
classical ``sampling'' algorithm.}
      \label{WorstCaseFigure}
\end{figure} 

      In figure~\ref{WorstCaseFigure} we
plotted the worst case probability of error against $\eta$ 
for both the quantum and the classical algorithm in the limit of large $N$. 
In the quantum case 
$\varepsilon$ decreases more rapidly and stays well below the classical
probability of error as long as $\eta$ is not too close to $1$ (remember
that the ``sampling'' algorithm is no longer probabilistic if we compute
our function over its entire domain!). 

      Looking now at the best case, we find that there is again a single 
class of functions which is easily dealt with by both algorithms, that
is one--to--one functions or permutations of the points in the
domain (this obviously requires $N$ to be equal to $M$). 

      Using the classical sampling algorithm, one can evidently be sure
to distinguish an invertible function from a constant one with only two
computations, since the former does not assume any value in its range
more than once.
      
      In the quantum case, permutations are associated with matrices 
having exactly one ``1'' in each row and in each column. Such matrices
turn out to be orthogonal to $\mathcal{C}^{\prime}$. 
Therefore, 
a measurement of the final state yielded by 
a permutation can either result in a FAIL
or in projection along $\overline{\mathcal{C}}$, which indicates
that the function is not constant.
Now FAILs can only 
be obtained with probability $1/N$, which luckily vanishes as $N$ grows
larger. We conclude that, in the limit of large $N$, the quantum algorithm
is practically guaranteed to spot a one--to--one function at first sight, 
after a single computation, thus doubling the efficiency of the
classical algorithm.

        By the way, we note that if we only had to tell constant functions
from permutations---if our practical problem didn't require us to
deal with non--invertible, non--constant functions---we would be back to
the original situation of Deutsch's example.
We can now see what was so special about the four functions considered
by Deutsch in his original example (see equation~\ref{TheFourFunctions}). 
When both the domain and the range consist of two points only all 
non--constant functions turn out to be one--to--one, so that all ambiguity
is removed.

\section{Numerical Results}

\begin{figure}
      \centering\epsfig{file=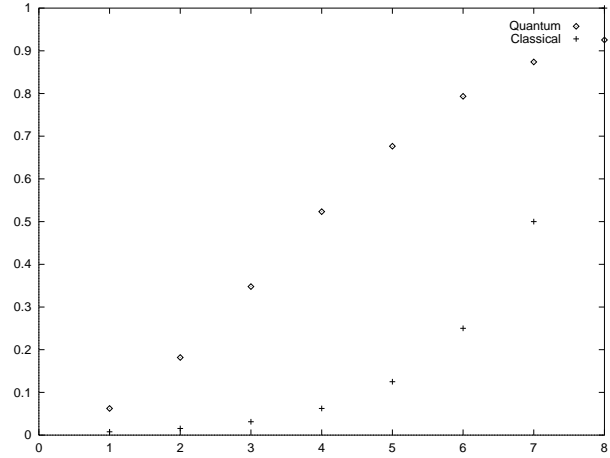, width=8.6cm}
      \caption{Posterior probability $\Pr(\mathrm{const}|k)$
versus $k$ ($N=8$; $M=2$; linear scale). 
Boxes: quantum algorithm;
crosses: classical ``sampling'' algorithm.}
      \label{Fig8x2Lin}
\end{figure}

\begin{figure}
      \centering\epsfig{file=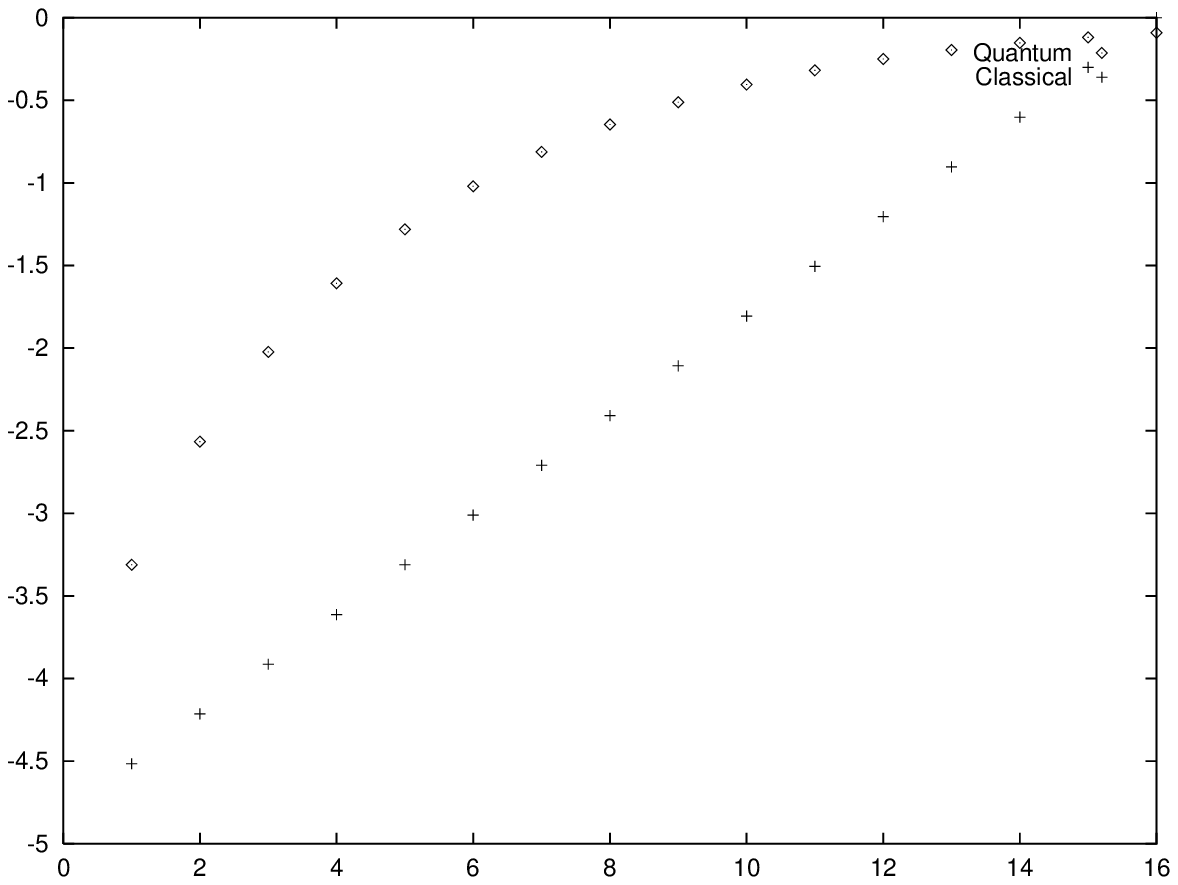, width=8.6cm}
      \caption{Posterior probability $\Pr(\mathrm{const}|k)$
versus $k$ ($N=16$; $M=2$; log~scale). 
Boxes: quantum algorithm;
crosses: classical ``sampling'' algorithm.}
      \label{Fig16x2Log}
\end{figure}

We are including, in figures~\ref{Fig8x2Lin} through~\ref{Fig24x24Log},
some comparative graphics of the posterior probabilities 
$\Pr(\mathrm{const}|k)$
expressed by equations~(\ref{bayes2}) and~(\ref{quantum}) versus 
the number $k$ of successful computations effected (by successful computation
we mean all computations barring FAIL results).

As our previous analysis suggested, the quantum
algorithm turns out to be far more efficient than the classical ``sampling''
algorithm for small values of $k/N$. We emphasize that this result is entirely
dependent upon the use of quantum parallelism. This
highly non--classical feature of quantum computation apparently allows a 
quicker exploration of the domain of function $f$, even in the case that the
investigated property is \emph{not} QPC.

The posterior probability $\Pr (\mathrm{const}|k)$ we used for our numerical
calculations is conditoned to a sequence of $k$ ``constant results'' of the
quantum algorithm. We have overlooked the possibility  of 
obtaining one or more FAIL outcomes. This is particularly significant when $M=2$
(figures~\ref{Fig8x2Lin} and~\ref{Fig16x2Log}), because in such cases a FAIL
result
has a $1/2$ probability to show up. This means that in order to obtain $k$
projections
of the final state of the computer along $\mathcal{C'}$ one must expect to
run the quantum computer $2k$  times. Nervertheless, as the graphics
show, the quantum strategy always turns out to be convenient, at least
for small values of $k/N$.

      We finally note that as $N$ and $M$ grow larger (see for instance
figure~\ref{Fig24x24Log}) the resulting posterior probabilities turn out
to be so low that both the quantum and the classical algorithm are 
virtually useless. This entirely depends on our assumption of an uniform
distribution over functions $f$, which is probably eccessively penalizing.
In real--world situations, we can expect
the quantum algorithm to be useful in any situation in which the ``sampling''
algorithm is successfully employed at the present day. 
  
\section{Acknowledgements}

We are grateful to C.\@~M.\@~Becchi for posing the question which led to this 
work. We acknowledge the interest of G.\@~Castagnoli and the collaboration
with Elsag--Bailey; we also thank A.\@~Ekert and C.\@~Macchiavello for  
interesting discussion. Special thanks to E.\@~Beltrametti for continuous 
help and advice.

\begin{figure}
      \centering\epsfig{file=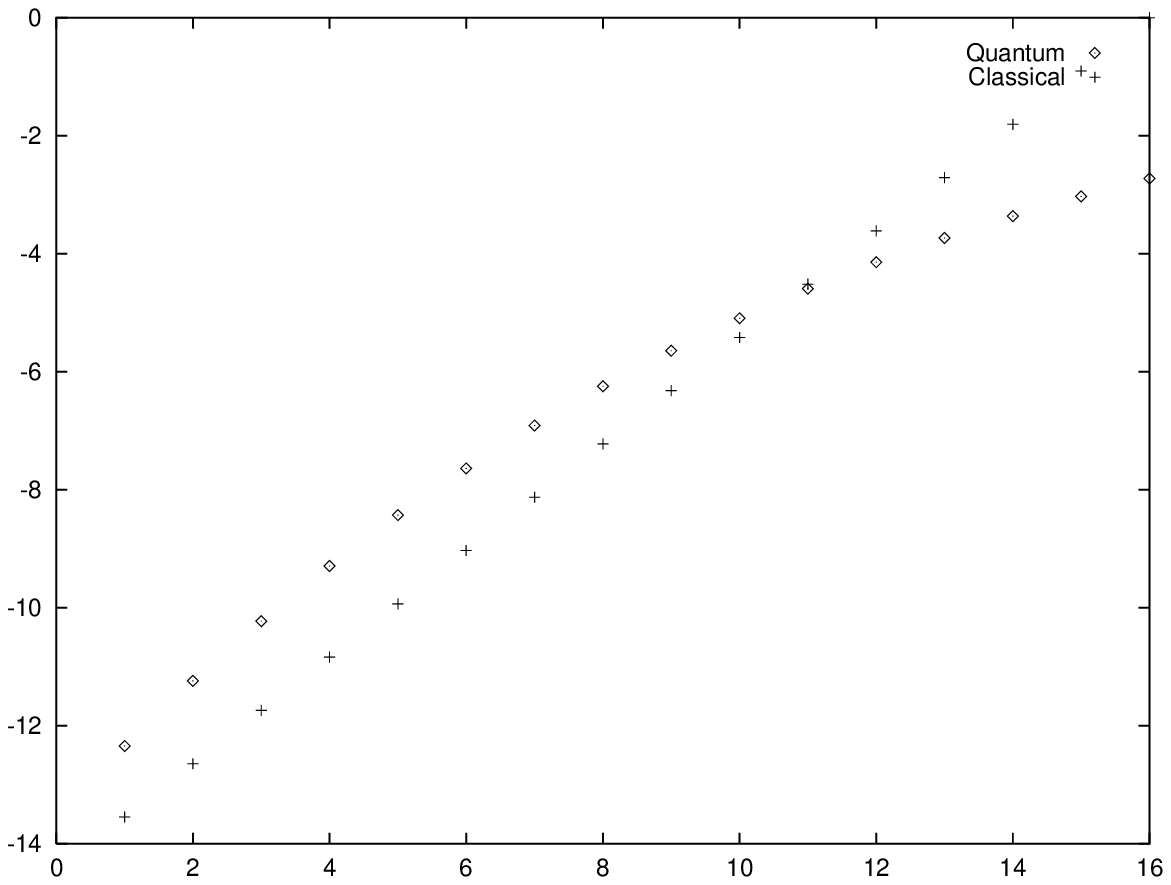, width=8.6cm}
      \caption{Posterior probability $\Pr(\mathrm{const}|k)$
versus $k$ ($N=16$; $M=8$; log~scale).
Boxes: quantum algorithm;
crosses: classical ``sampling'' algorithm.}
\end{figure}

\begin{figure}
      \centering\epsfig{file=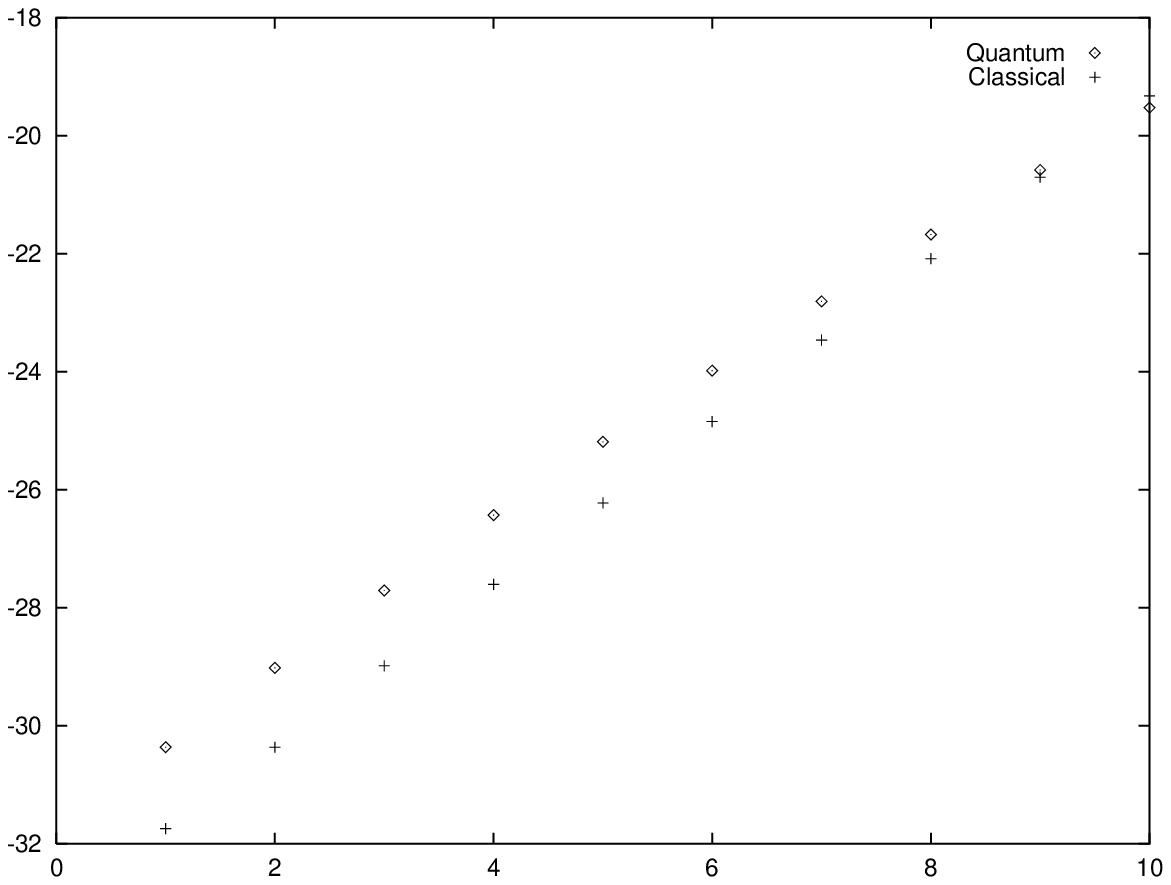, width=8.6cm}
      \caption{Posterior probability $\Pr(\mathrm{const}|k)$
versus $k$ ($N=24$; $M=24$; log~scale).
Boxes: quantum algorithm;
crosses: classical ``sampling'' algorithm.}
      \label{Fig24x24Log}
\end{figure}
\end{document}